# Propriétés des chenaux optiques, leur chromatisme et ses effets sur la réfraction astronomique

*Properties of optical ducts, their chromatism and its effects on astronomical refraction*

**Luc Dettwiller** [a]

[a] *Université Jean Monnet Saint-Etienne, CNRS, Institut d Optique Graduate School, Laboratoire Hubert Curien UMR 5516, F-42023, SAINT-ETIENNE, France*
E-mail: dettwiller.luc@gmail.com

**Abstract.** The fundamental quadrature governing light rays in a spherically symmetrical medium is first recalled. A rigorous discussion of some qualitative properties of its solutions follows, using the Young-Kattawar diagram which leads to a geometric formulation of the ray curvature. The case of an optical duct is deepened, analyzing transfer curves for different positions of the observer with respect to the duct. New analytical expressions for their wavelength dependence are derived, and their numerical consequences are coherent with computer simulations.

**Résumé.** Après avoir rappelé la quadrature fondamentale régissant les rayons lumineux dans un milieu à symétrie sphérique, on montre une discussion rigoureuse de propriétés qualitatives de ses solutions en utilisant le diagramme d'Young-Kattawar, et on donne avec lui une formulation géométrique de la courbure des rayons. On s'appesantit sur le cas d'un chenal optique, et des courbes de transfert (pour différentes positions de l'observateur par rapport à lui), ainsi que sur des expressions inédites de leur variation avec la longueur d'onde, donnant lieu à des applications numériques en accord avec des résultats tirés de simulations informatiques.

**Mots clés.** Réfraction astronomique, courbure des rayons, théorème de Biot, grandissement angulaire vertical local, accourcissement, théorème de Biot-Sang-Meyer-Fraser-White, chenal optique, courbe de transfert, liseré vert, liseré rouge, dispersion chromatique d'un chenal.

## 1. Introduction

Il est bien connu que dans un modèle d'atmosphère terrestre à symétrie sphérique, d'indice de réfraction $n(r)$ – où $r$ désigne la distance au centre $C$ de la Terre (voir la Figure 1) – la forme des rayons lumineux (R) est caractérisée par l'invariant de Bouguer

$$I_B := n\, r \sin\alpha \tag{1}$$

où $\alpha$ est l'angle entre (R) et le rayon vecteur (le symbole $:=$ indiquant une définition ou une notation).

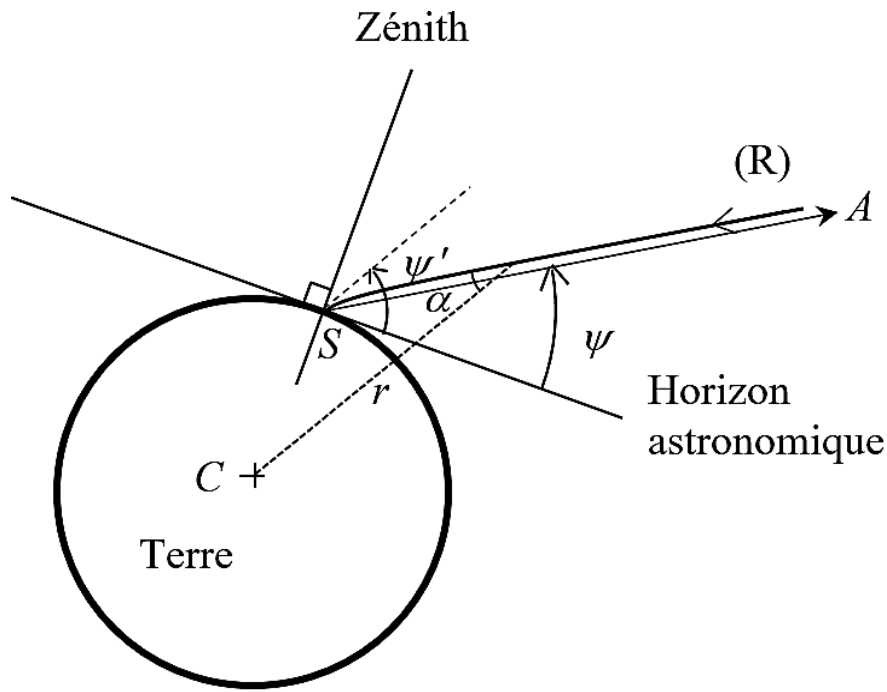

**Figure 1.** Notations utiles pour l'étude des rayons lumineux (R) arrivant sur l'observateur $S$ dans l'atmosphère terrestre supposée à symétrie sphérique.

On sait en déduire, sous la forme d'une quadrature, l'équation différentielle régissant $r(\theta)$ qui caractérise (R) en coordonnées polaires ; nous rappellerons d'abord comment l'obtenir. Ensuite, parmi ses multiples applications, nous nous intéresserons à l'étude des rayons guidés dans l'atmosphère terrestre, où la distribution d'indice forme parfois un chenal optique, en un sens précis que nous redonnerons ; nous nous appuierons sur des méthodes graphiques permettant de déduire beaucoup de propriétés de (R) qualitatives mais rigoureuses, à partir du diagramme d'Young-Kattawar qui donne le graphe de $f(r) := r\,n(r)$, et nous donnerons grâce à lui une formulation géométrique de la courbure de (R). Enfin, nous nous intéresserons aux propriétés optiques des chenaux, traduites par les courbes de transfert (donnant la hauteur angulaire vraie $\psi$ d'un astre $A$ en fonction de sa hauteur apparente $\psi'$) ; nous étudierons surtout leur variation avec la longueur d'onde $\lambda_0$ dans le vide (paramétrant les courbes), et nous présenterons sur ce sujet des calculs analytiques inédits à notre connaissance, dont nous tirerons des valeurs numériques concrètes.

## 2. Quadrature

La planéité de (R) dans un milieu à symétrie sphérique permet d'utiliser les coordonnées polaires $(r, \theta)$, avec lesquelles on a

$$\frac{1}{\tan\alpha} = \frac{1}{r}\frac{dr}{d\theta}, \tag{2}$$

et

$$1 + \frac{1}{\tan^2\alpha} = \frac{1}{\sin^2\alpha}, \tag{3}$$

dont on déduit l'équation différentielle du rayon sous la forme d'une quadrature [1] :

$$\left(\frac{dr}{d\theta}\right)^2 = r^2\left[\frac{f^2(r)}{I_B^2} - 1\right] := h(r, I_B) \quad \text{avec} \quad I_B = f(r_S)\cos\psi'. \tag{4}$$

Concernant les solutions, le diagramme d'Young-Kattawar [2], *i.e.* le graphe de $f(r)$ avec la droite d'ordonnée $I_B$ (voir la Figure 2) permet des discussions graphiques plus simples que par le graphe de $h$ à $I_B$ fixé. Notons que $I_B$ fixe le rayon, à une rotation près autour de $C$. Cette quadrature (4) a de multiples applications ; on peut en voir une simple au sous-paragraphe 3.6 de notre article [3], une plus évoluée dans [4] (ces deux textes faisant partie du numéro spécial des C. R. Phys. intitulé *Astronomie, atmosphères et réfraction*), et ci-dessous on en montre d'autres, sans prétendre à l'exhaustivité.

## 3. Discussion graphique des rayons sur un graphe équivalent à celui de $h$ ; cas d'un chenal optique

À titre d'exemple instructif, considérons le cas où $f(r) - I_B$ est strictement positif sur l'intervalle $]r_1(I_B), r_2(I_B)[$ aux bornes (non nulles) duquel il présente des annulations simples, *i.e.* où la dérivée $f'\left(r_{1,2}(I_B)\right)$ de $f(r) - I_B$ est non nulle (Figure 2) ; c'est le cas aussi pour $h$ à $I_B$ fixé, puisque

$$\frac{\partial h}{\partial r}\left(r_{1,2}(I_B), I_B\right) = 2\,r_{1,2}^2(I_B)\,f'\left(r_{1,2}(I_B)\right)\big/I_B \quad . \tag{5}$$

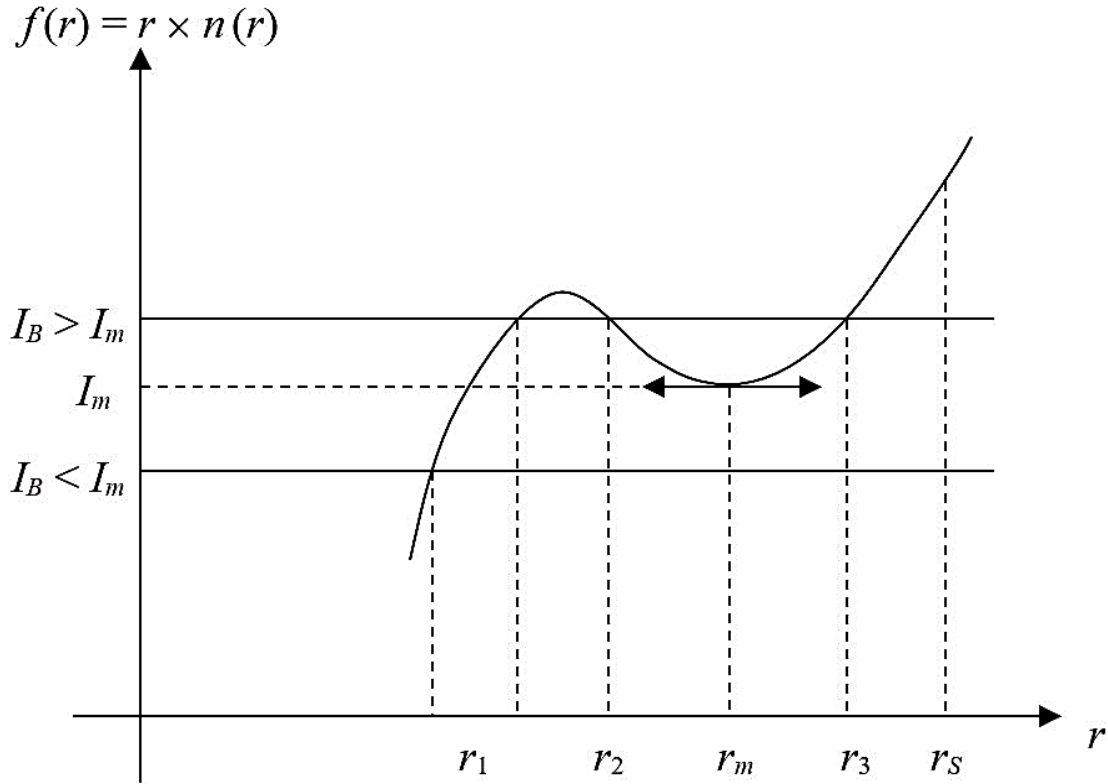

**Figure 2.** Diagramme d'Young-Kattawar.

Alors les rayons lumineux correspondants oscillent entre deux cercles de centre $C$ et de rayons $r_1(I_B)$, $r_2(I_B)$. Mais ils ne présentent pas de point méplat ni d'inflexion (leur concavité restant toujours tournée vers $C$) tant que la fonction $r \mapsto n(r)$ reste monotone, ce qui (dans l'atmosphère sous nos latitudes) exige $dT/dr \neq -34\ \mathrm{K \cdot km^{-1}}$ indépendamment de la longueur d'onde, la température et la pression (voir le sous-paragraphe 7.1.4 de notre article [5]) – mais ceci n'est plus vrai dans l'ionosphère pour le domaine radio. Aux péricentres des rayons lumineux comme en leurs apocentres, ils subissent une quasi-réflexion au sens d'Everett [6]. Dans le cas limite où $r_1 = r_2 = r_C$ forme une annulation double de $f(r) - I_B$, on a $f'(r_C) = 0$, $\partial h/\partial r(r_C, I_B) = 0$, et on trouve la condition de trajet circulaire de rayon $r_C$ ; on voit que ce trajet est stable si $f''(r_C) < 0$, *i.e.* $\partial^2 h/\partial r^2(r_C, I_B) < 0$. Lorsque pour $r_m$ et $r_m'$ la fonction continue $f$ prend la même valeur sans le faire pour les $r$ intermédiaires, et que $f(r_m)$ est un minimum stationnaire de $f$ noté $I_m$ (voir la Figure 2), alors la coquille sphérique de centre $C$ et de rayons $r_m$ et $r_m'$ est un chenal optique, *i.e.* un domaine connexe ($C$) qui est le plus grand domaine ($\mathcal{D}$) contenant tous les autres tels que, pour tout point $M_0$ de l'intérieur de ($\mathcal{D}$), il existe un ensemble ($\mathcal{E}_0$) de directions de mesure non nulle tel que chaque rayon lumineux passant par $M_0$ avec une des directions de ($\mathcal{E}_0$) reste dans ($\mathcal{D}$) (en omettant les effets de bords) ; le domaine ($C$) est aussi appelé « guide d'ondes », et on dit que les rayons qui restent dans ($C$) sont guidés. Les cercles de centre $C$ et de rayon $r_m$ sont des trajets instables pour la lumière, et si $f'(r_m') \neq 0$ les cercles de centre $C$ et de rayon $r_m'$ ne sont pas du tout des trajets lumineux possibles. Chaque axe apsidal, *i.e.* passant par $C$ ainsi que par un péricentre ou un apocentre du rayon, est un axe de symétrie pour celui-ci. L'angle entre deux axes apsidaux consécutifs s'écrit $\int\limits_{r_1(I_B)}^{r_2(I_B)} \dfrac{dr}{\sqrt{h(r, I_B)}}$ ; le rayon lumineux est invariant par toute rotation d'un multiple pair de cet angle, autour de $C$, dans son plan.

Sur ce diagramme, on peut aussi traduire graphiquement la courbure d'un rayon [7,8], en choisissant judicieusement l'orientation arbitraire des angles de façon que $\alpha \in [0, \pi]$ :

$$\frac{1}{\mathcal{R}} = \frac{\boldsymbol{n} \cdot \boldsymbol{grad}\, n}{n} = \frac{\sin \alpha}{n} \left| \frac{dn}{dr} \right| = \frac{I_B\, r}{f^2(r)} \left| \frac{df}{dr} - \frac{f}{r} \right| \frac{1}{r} = \frac{\mathrm{HJ}}{\mathrm{HP}^2} \frac{\mathrm{OL}}{\mathrm{OH}} \tag{6}$$

où

$$\overline{\mathrm{OL}} = f - r \frac{df}{dr} = -r^2 \frac{dn}{dr}. \tag{7}$$

Si $\overline{\mathrm{OL}} > 0$, la concavité est tournée vers $C$, et $dn/dr < 0$ (le contraire sinon) ; le rayon s'infléchit si L traverse O, ce qui veut dire que $dn/dr$ s'annule et change de signe, comme $dT/dr + 34 \ \mathrm{K} \cdot \mathrm{km}^{-1}$ dans l'atmosphère terrestre [5, sous-paragraphe 7.1.4].

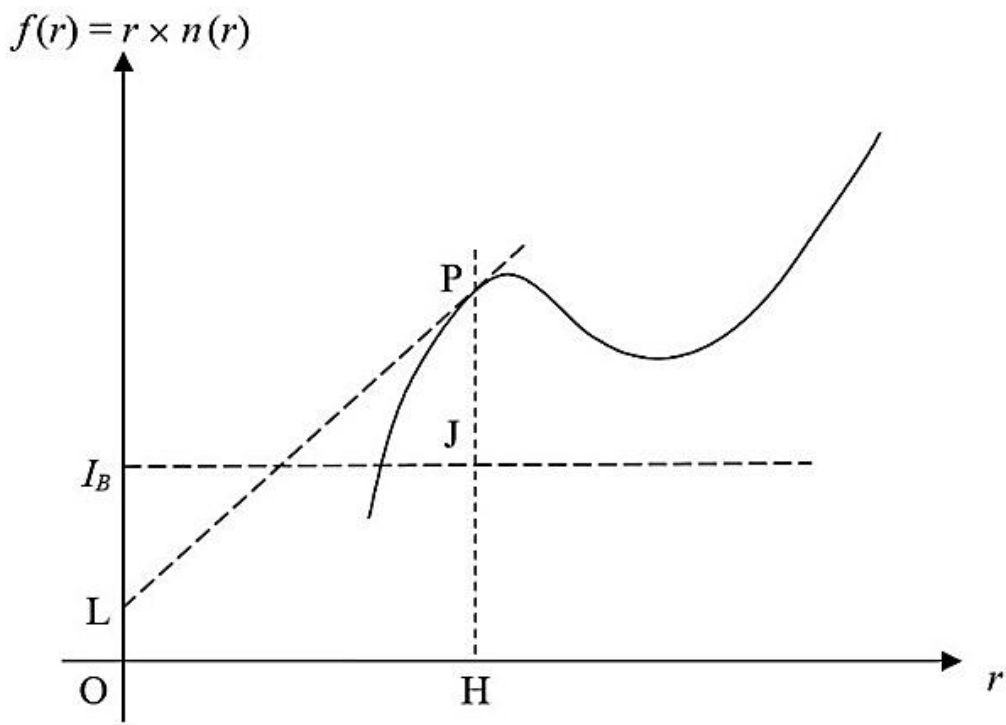

**Figure 3.** Détermination de la courbure d'un rayon à l'aide du diagramme d'Young-Kattawar : $\dfrac{1}{\mathcal{R}} = \dfrac{\mathrm{HJ}}{\mathrm{HP}^2} \dfrac{\mathrm{OL}}{\mathrm{OH}}$ .

### 3.1. Exemple de réalisation

Un cas réaliste de diagramme du type de la Figure 2 est produit par une couche d'inversion allant de 50 m à 60 m d'altitude, d'amplitude 2 °C donc modérée (voir la Figure 4) ; cependant nous dirons qu'elle est « raide », car son gradient thermique est $200 \ \mathrm{K} \cdot \mathrm{km}^{-1}$, supérieur au gradient minimum pour le guidage ($116 \ \mathrm{K} \cdot \mathrm{km}^{-1}$).

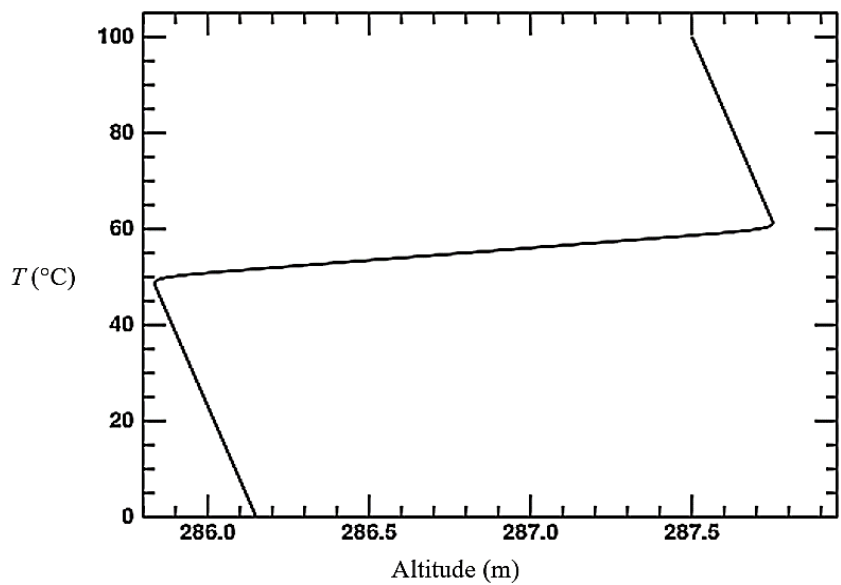

**Figure 4.** Profil de température $T$ caractérisant la couche d'inversion raide qui produit le chenal donnant lieu aux phénomènes montrés par les Figures 5-10. Ce graphe traduit un modèle simpliste occultant les détails du profil thermique réel, mais ceux-ci ont très peu d'influence sur les phénomènes optiques vus par un observateur situé hors du chenal.
© Andrew T. Young, https://aty.sdsu.edu/explain/simulations/ducting/duct_intro.html

Alors le bas du chenal est entre 45 m et 46 m, variant avec $\lambda_0$ – voir la Figure 5, qui aide à comprendre que le chenal est occupé par la couche d'inversion dans sa partie haute, mais qu'il a aussi une partie supplémentaire sous la couche d'inversion, à cause de la symétrie sphérique.

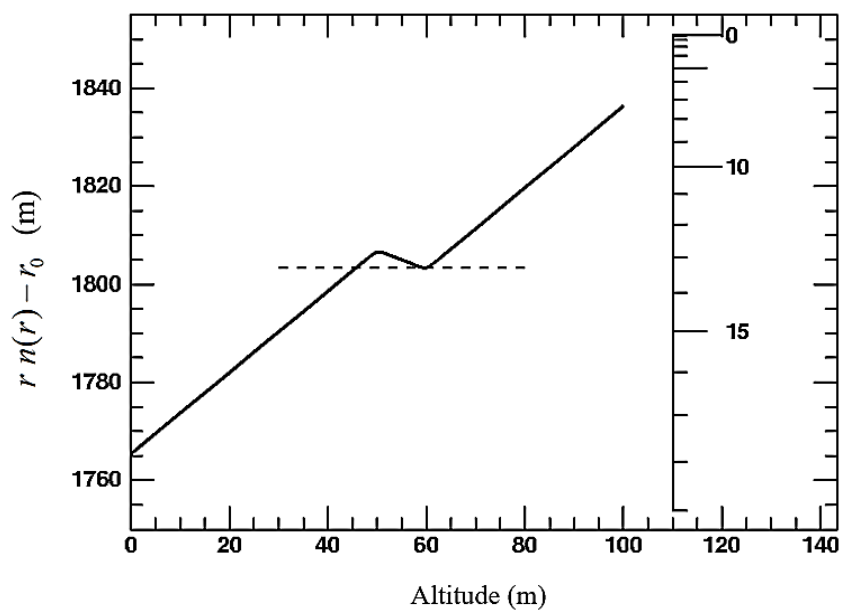

**Figure 5.** Diagramme d'Young-Kattawar indiquant les bornes du chenal, pour une valeur donnée de $\lambda_0$. Un décalage est opéré dans les échelles des abscisses et des ordonnées, qui indiquent respectivement (en mètres) $r - r_0$ et $r\,n(r) - r_0$ où $r_0$ désigne le rayon

terrestre. L'échelle verticale irrégulière à droite de la courbe indique de combien il faut descendre la droite horizontale correspondant à $I_B$ quand $|\psi'|$ passe de $0'$ à $19'$ au maximum. Les abscisses des intersections de la ligne pointillée avec la courbe en trait plein sont les altitudes des frontières basse (45 à 46 m) et haute (60 m) du chenal – cette dernière étant quasiment l'altitude du sommet de la couche d'inversion, car le domaine d'altitudes de la portion décroissante de la courbe correspond presque à celui de la couche d'inversion.
© Andrew T. Young, https://aty.sdsu.edu/explain/atmos_refr/dip_diag.html

Dans le présent article, restant dans l'étude de la réfraction astronomique en présence de la couche d'inversion raide, nous montrons maintenant les différentes courbes de transfert (notion définie au paragraphe 1) en fonction de la position de $S$. Trois cas se présentent ; avec le premier d'entre eux à titre d'exemple, nous illustrerons son intérêt pour étudier la déformation de l'image d'un objet extra-atmosphérique étendu (comme le Soleil ou la Lune) due à la réfraction.

### 3.2. Courbe de transfert pour un observateur au-dessus du chenal

On considère le cas où $r_S > r_m(\lambda_0)$, comme sur la Figure 2 ; donc $f(r_S) > I_m := f(r_m)$ – nous omettons ici d'écrire la variable $\lambda_0$, car nous raisonnons ici à $\lambda_0$ fixé. Comme $I_B = n_S \, r_S \cos\psi'$ diminue quand $|\psi'|$ augmente, on voit que pour $\psi'$ égal à

$$\psi'_a := -\arccos\frac{f(r_m)}{f(r_S)} \tag{8}$$

la valeur de $r$ au péricentre des rayons présente une discontinuité en fonction de $\psi'$. Le point important à retenir est que, dans la situation où le minimum local de $f$ est stationnaire (ce qui est le cas réaliste), alors la courbe de transfert présente une asymptote bilatérale verticale d'abscisse $\psi'_a$ (voir la Figure 6) ; cela vient de la divergence de $\int_{r_m}^{r_S} dr/\sqrt{h(r, I_m)} = +\infty$. Notre commentaire [4] d'un article de Kummer [9] rappelle que celui-ci fait grand cas du caractère impropre de $\int_{r_m}^{r_S} dr/\sqrt{h(r, I_m)}$, et nous y montrons alors en appendice que $\psi(\psi')$ est équivalent, au voisinage de $\psi'_a$, à $\ln|\psi' - \psi'_a|$ multiplié par une constante, mais que celle-ci est à gauche le double de la constante à droite, contrairement aux apparences.

De la Figure 6 on peut déduire la séquence des images du Soleil pendant un coucher. L'objet y est représenté, sur l'axe des ordonnées, par un segment $[\psi_d, \psi_u]$ dont le milieu est la hauteur angulaire vraie $\psi_0$ du centre de Soleil, et la largeur $\psi_u - \psi_d \cong 32'$. Au fil du coucher, $\psi_0$ décroît, et les images sont déduites du correspondant de ce segment sur l'axe des abscisses d'après le graphique de la Figure 6. Un phénomène approchant de ce cas théorique a été filmé [10] par Jean-Luc Dauvergne depuis le Pic du Midi, assez haut pour se trouver souvent au-dessus d'une couche d'inversion.

Sur la Figure 6 on voit que pour $\psi_d > -35'$ l'image du Soleil est ordinaire, avec un accourcissement dû au grandissement angulaire vertical $d\psi'/d\psi \cong 0,8$ conformément au théorème de Biot (voir le sous-paragraphe 7.1 de [5]).

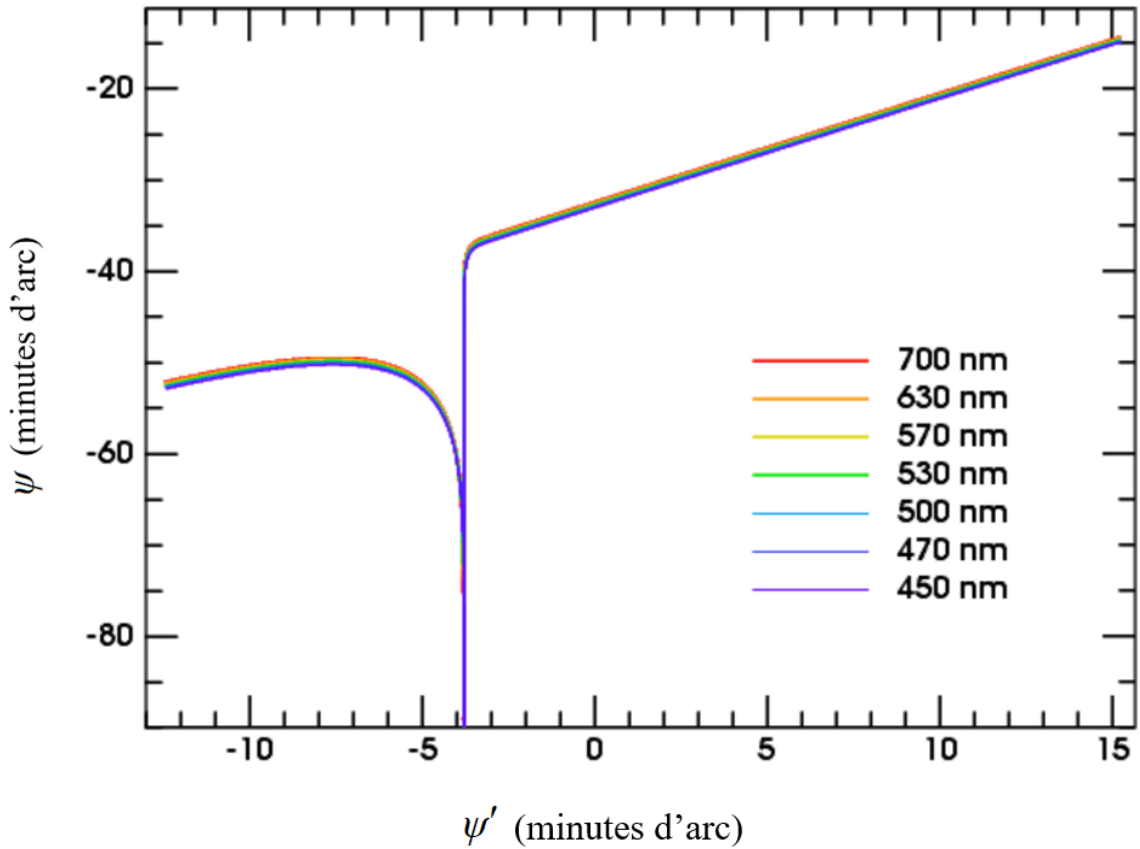

**Figure 6.** Courbes de transfert pour un observateur $S$ dont l'élévation est $E = 65$ m (donc $S$ est 5 m au-dessus du chenal), à différentes longueurs d'onde – allant de 450 nm à 700 nm. Leur domaine de définition admet une borne inférieure qui est ici vers – 14′ (en fait, c'est l'opposé de la dépression $\delta$ de l'horizon). La pente sur l'axe des ordonnées est fixée par le théorème de Biot [5]. L'abscisse de l'asymptote bilatérale, toujours négative (conformément au théorème de Biot-Sang-Meyer-Fraser-White [5]), tend vers 0 avec la distance entre $S$ et la frontière haute du chenal. La position de celle-ci varie beaucoup moins (du rouge au bleu) que celle de la frontière basse ; les courbes pour les différentes longueurs d'onde sont beaucoup plus serrées que sur les Figures 9-10.
© Andrew T. Young, https://aty.sdsu.edu/explain/simulations/transfer/transfer.html

Sur la Figure 7 de gauche on a $\psi_d$ très légèrement inférieur au maximum local des courbes de transfert $-50'$, et cela d'autant plus pour la courbe rouge ; ainsi s'explique la « bulle du Soleil couchant », frangée de rouge, qui est vue sous l'image droite supérieure, et constituée de la jonction d'une image droite d'un fragment bas du Soleil avec l'image renversée de ce même fragment. Pour une longueur d'onde donnée, certains points du Soleil ont donc trois images, alors pour elle l'image du Soleil est formée de trois composantes, qui sont, de haut en bas, respectivement droite, renversée et droite. Le bord quasi rectiligne du bas de l'image droite supérieure est dû à une portion des courbes de transfert très proche de l'asymptote verticale, où $d\psi/d\psi' \gg 1$, ce qui donne donc d'une partie du Soleil une image droite très aplatie car $d\psi'/d\psi \ll 1$, mais l'image droite supérieure reste quand même celle du Soleil entier. L'espace vide entre la 1ᵉ et la 2ᵉ composante de cette image pour une longueur d'onde donnée correspond à l'écart horizontal, à l'ordonnée $\psi_d$, entre la branche décroissante et la branche croissante de part et d'autre de l'asymptote bilatérale verticale pour cette longueur d'onde. Il s'amincit au fur et à mesure que le Soleil se couche, et il devient vite imperceptible, quoiqu'il ne disparaisse jamais en théorie, vu la présence de l'asymptote bilatérale (pour la courbe de transfert de cette longueur d'onde) ; à cause de celle-ci, l'image droite supérieure et l'image renversée ne se couchent jamais en théorie, elles existent toujours et restent toujours (malgré leur aplatissement croissant) des images du Soleil entier, mais le flux lumineux arrivant sur ces images (rétiniennes par exemple) devient de plus en plus faible, d'une part parce que leur aire tend vers zéro puisque leur aplatissement tend vers l'infini, d'autre part parce que leur luminance tend aussi vers zéro – à cause de l'extinction croissante due à l'allongement du trajet des rayons lumineux dans l'atmosphère. Seule l'image droite inférieure (*i.e.* la 3$^e$ composante) se couche vraiment : elle cesse d'exister dès que $\psi_u < \psi(-\delta)$ où $\delta$ désigne la dépression de l'horizon, *i.e.* l'angle mesurant, pour $S$ ayant une certaine élévation au-dessus de la mer, la « *bassesse apparente de l'horizon de la Mer* » [11] par rapport à l'horizontale – cette grandeur joue un rôle très important en navigation astronomique [3,4].

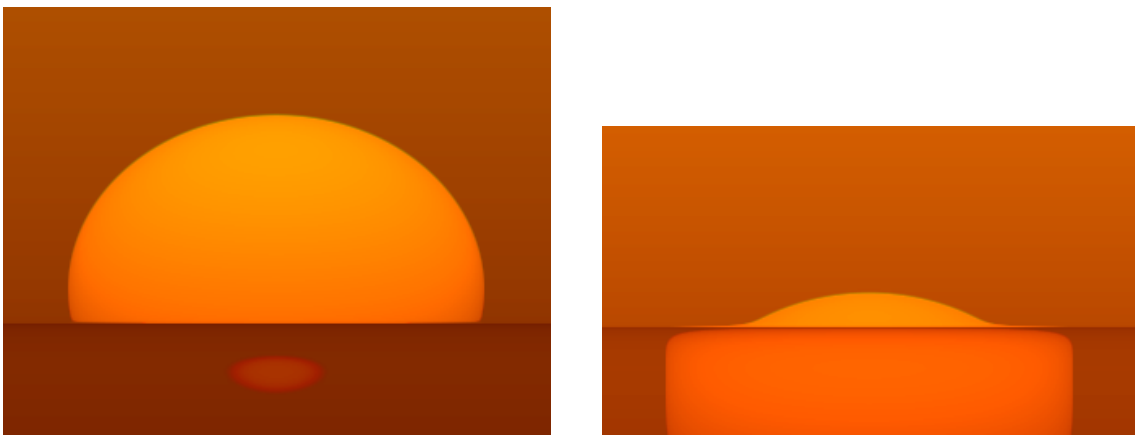

**Figure 7.** Image du Soleil vue de 65 m d'élévation ; de gauche à droite, la hauteur angulaire vraie du centre du Soleil passe de –34′ à –50′. Sur la Figure de gauche, noter un discret liseré vert en haut de l'image droite supérieure, et un liseré rouge entourant complètement la portion d'image apparue en-dessous de l'image droite supérieure. Sur la Figure de droite, noter les pointes horizontales de l'image droite supérieure, corroborant le fait que celle-ci est l'image du Soleil entier, mais avec un très fort aplatissement pour la majeure partie de celui-ci.
© Andrew T. Young, https://aty.sdsu.edu/explain/simulations/ducting/duct_intro.html

Sur la Figure 7 de droite, on a $\psi_0$ quasiment égal au maximum local des courbes de transfert (i.e. à –50′.), où $d\psi/d\psi' = 0$, donnant donc une composante d'image droite et une renversée, mais très étirées et qui se joignent là où le grandissement angulaire vertical local est infini.

Sur la Figure 8 on voit la suite du coucher de Soleil, avec $\psi_0$ de plus en plus inférieur au maximum local des courbes de transfert. L'image droite supérieure, que l'on voyait majoritairement en haut sur la Figure 7 de gauche, est devenue tellement comprimée verticalement qu'elle est imperceptible sur la Figure 8, où de haut en bas de chaque image on ne voit plus que la 2$^e$ et la 3$^e$ composante (respectivement renversée et droite) de l'image du Soleil, qui se séparent quand $\psi_u$ passe en dessous de $-50'$.

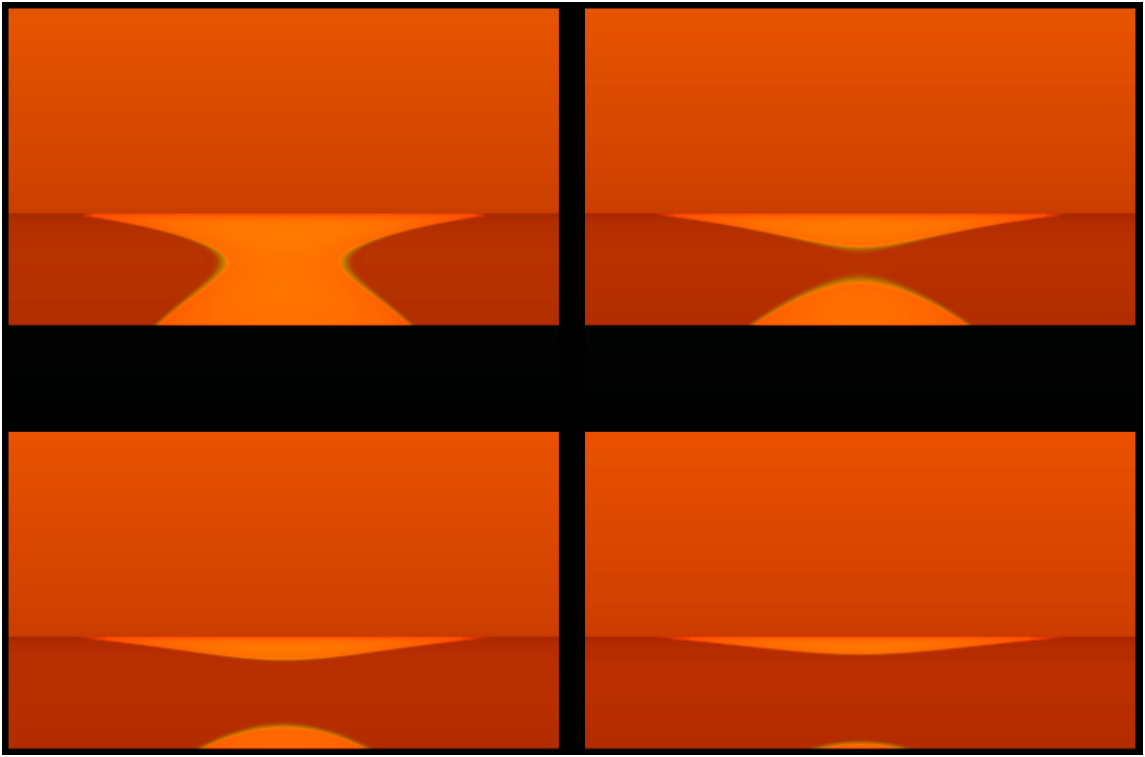

**Figure 8.** Image du Soleil vue de 65 m d'élévation ; dans le sens usuel de lecture, la hauteur angulaire vraie $\psi_0$ du centre du Soleil vaut successivement $-65'$ puis $-66'$, $-67'$ et $-68'$. On ne voit concrètement qu'une des deux portions droites de l'image (c'est la portion inférieure, due à la partie croissante de la courbe de transfert à gauche de l'asymptote), surmontée d'une portion d'image renversée (« pseudo-mirage »), due à la partie décroissante de la courbe de transfert ; l'autre portion droite, due à la partie de la courbe de transfert à droite de l'asymptote, est tellement comprimée verticalement qu'elle est imperceptible.
© Andrew T. Young, https://aty.sdsu.edu/explain/simulations/ducting/duct_intro.html

Une image renversée est due à la partie décroissante de la courbe de transfert ; on l'appelle « pseudo-mirage » [12]. Le préfixe *pseudo* signale ici une particularité de ce cas de mirage, partagée aussi par le *Nachspiegelung* : les rayons formant l'image renversée ne subissent pas de quasi-réflexion *totale* (au sens d'Everett – voir le sous-paragraphe 2.1 de [13]), contrairement à ce qui se passe avec les mirages inférieurs. Un observateur au-dessus de la couche d'inversion peut voir un pseudo-mirage même si celle-ci n'est pas raide ; ce dernier cas est beaucoup plus fréquent que celui du mirage supérieur ou du *Nachspiegelung*, où l'image renversée est due à une couche d'inversion raide.

## 4. Courbe de transfert pour un observateur dans le chenal

Il apparaît maintenant deux asymptotes unilatérales verticales, symétriques par rapport à l'axe des ordonnées, car elles correspondent aux hauteurs angulaires apparentes à $\pm\,\psi'_a(r_S,\lambda_0) = \mp \arccos \dfrac{f\left(r_m(\lambda_0),\lambda_0\right)}{f\left(r_S,\lambda_0\right)}$ où on utilise la fonction $f$ généralisée, définie par $f(r,\lambda_0) := r\,n(r,\lambda_0)$. Entre ces valeurs la fonction n'est plus définie (voir la Figure 9) ; l'écart entre les asymptotes unilatérales traduit la bande vide de Wegener [14] où la fonction de transfert n'est pas définie, ce qui veut dire qu'on ne peut y voir aucune image d'un quelconque objet extra-atmosphérique. Cet écart entre asymptotes est critique en $r_S = r'_m(\lambda_0)$, mais pas en $r_S = r_m(\lambda_0)$ ; cela signifie que le graphe de l'abscisse des asymptotes en fonction de $r_S$ présente une demi-tangente verticale en $r_S = r'_m(\lambda_0)$, à gauche de laquelle il n'est pas défini – ce graphe est en fait un diagramme de bifurcation, sans demi-tangente verticale au niveau de son point triple $\left(r_S = r_m(\lambda_0), \psi'_a(r_m(\lambda_0), \lambda_0) = 0\right)$.

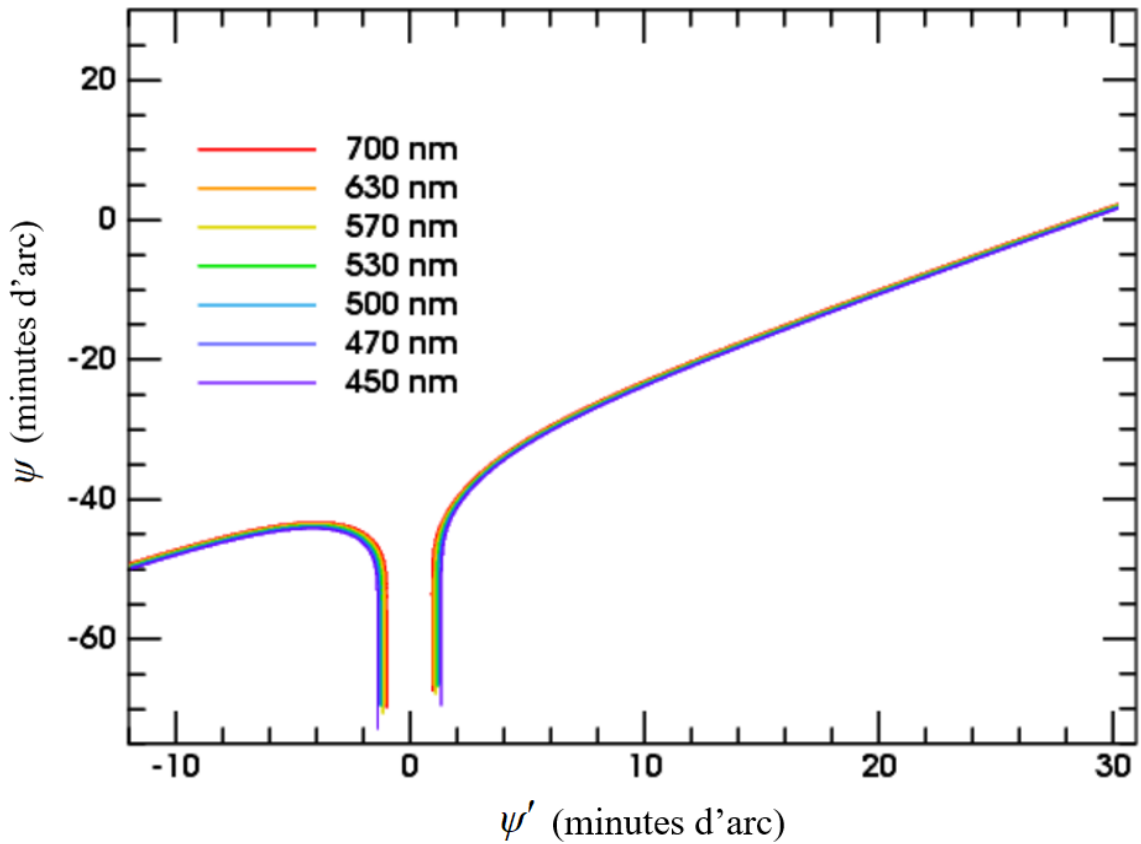

**Figure 9.** Courbes de transfert pour un observateur $S$ dans un chenal (à peine 1 m au-dessus de sa frontière basse), à différentes longueurs d'onde (allant de 450 nm à 700 nm). On observe que les asymptotes unilatérales sont symétriques par rapport à l'axe des ordonnées, et que plus l'air est réfringent, plus elles sont écartées. Comme leur écart varie critiquement en fonction de la distance de $S$ à la frontière basse du chenal, il est bien visible sur cette Figure – ainsi que sa variation avec la longueur d'onde.
© Andrew T. Young, https://aty.sdsu.edu/explain/simulations/transfer/transfer.html

Noter une partie décroissante de la courbe de transfert ; elle n'existe que sous la bande vide. C'est le *Nachspiegelung*, prédit par Wegener en 1918 en même temps que la bande vide, mais sous une forme simplifiée (car Wegener adoptait un modèle d'atmosphère avec une couche inférieure homogène, séparée de la couche supérieure d'indice plus faible par un dioptre sphérique) et sous une forme tronquée (la discontinuité d'indice dans le modèle de Wegener induit une courbe de transfert sans asymptotes verticales, mais avec des demi-tangentes verticales cependant aux bords de la bande vide) [14]. Quoiqu'ils donnent une image renversée, les rayons lumineux qui la forment ne subissent aucune quasi-réflexion, comme le montre facilement le diagramme d'Young-Kattawar.

## 5. Courbe de transfert pour un observateur sous le chenal

Dans ce cas la courbe de transfert n'a plus d'asymptote (voir la Figure 10), elle a un minimum critique aussi en $r_S = r'_m(\lambda_0)$, au sens où il tend vers $-\infty$ si l'observateur $S$ sous le chenal tend vers la frontière basse du chenal.

Il y a encore un *Nachspiegelung*, correspondant à la partie décroissante de la courbe de transfert, qui subsiste parce que l'observateur sous le chenal est encore près de celui-ci ; mais ce phénomène est plus facile à voir quand $S$ est au-dessus de la couche d'inversion – soit faible (de 0,8 K sur 20 m d'épaisseur typiquement, qui ne forme pas un chenal), soit modérée (avec 2 K sur 10 m par exemple) mais raide.

On constate que ces courbes de transfert, comme toutes les précédentes, corroborent le théorème de Biot-Sang-Meyer-Fraser-White (voir le sous-paragraphe 7.2 de [5]) : $\forall \psi' \geq 0 \quad d\psi'/d\psi \geq 0$.

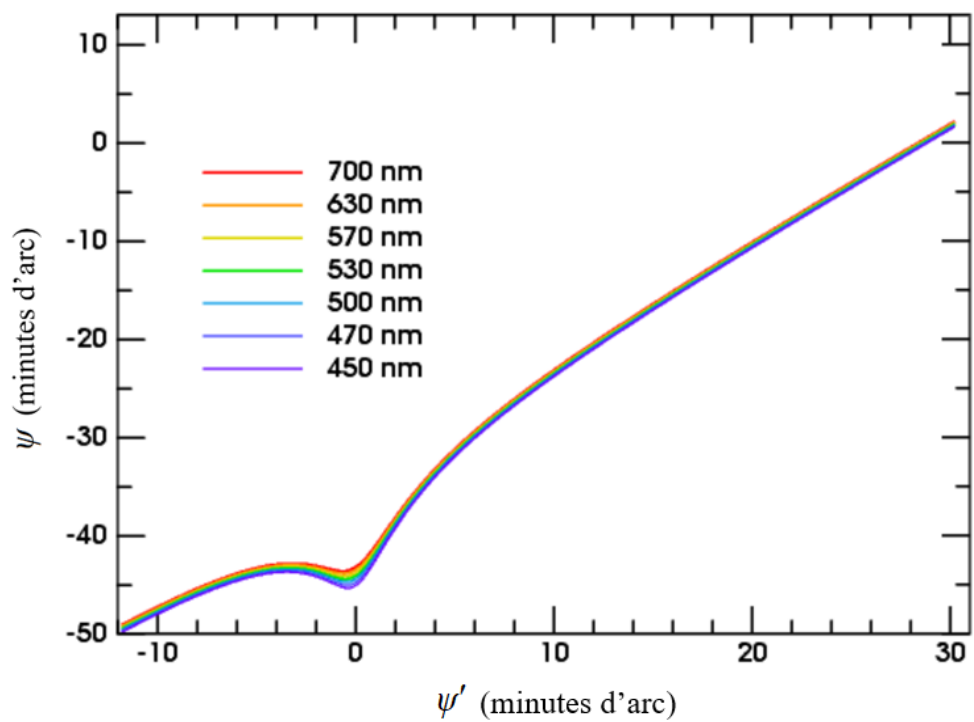

**Figure 10.** Courbes de transfert pour un observateur juste sous un chenal, à différentes longueurs d'onde (allant de 450 nm à 700 nm). Le minimum est d'autant plus accusé que l'air est plus réfringent. Son abscisse, toujours négative (conformément au théorème de Biot-Sang-Meyer-Fraser-White), tend vers 0 avec la distance entre $S$ et la frontière basse du chenal ; son ordonnée est critique aussi sur cette frontière, au sens où elle tend vers moins l'infini quand $r_S$ tend vers $r'_m(\lambda_0)_-$, et c'est pour cela que sa variation du rouge au bleu est importante, comparée à la variation d'ordonnée pour $r_S$ assez éloigné de $r'_m(\lambda_0)$. L'exaltation de la dispersion qui en résulte explique la formation d'une forme très spectaculaire de rayon vert.
© Andrew T. Young, https://aty.sdsu.edu/explain/simulations/transfer/transfer.html

Comme les changements de ces courbes sont critiques lorsque l'observateur traverse la frontière basse du chenal, il est important de savoir comment celle-ci se déplace en fonction de la longueur d'onde ; nous verrons que l'effet de son déplacement est beaucoup plus important que celui de la frontière haute, et que celui des asymptotes de la courbe de transfert.

**6. Dispersion des asymptotes verticales de la courbe de transfert**

Que l'observateur soit dans le chenal ou au-dessus, toute asymptote correspond à une des abscisses $\pm\,\psi'_a(\lambda_0)$ telles que

$$\psi'_a(\lambda_0) = -\arccos\left[\, f\left(r_m(\lambda_0),\lambda_0\right)/f\left(r_S,\lambda_0\right)\right] \quad \forall\,\lambda_0 \tag{9}$$

La dérivation logarithmique de $\cos\psi'_a(\lambda_0)$, en tenant compte de la stationnarité de $f$ en $r_m$ et de la loi empirique de Gladstone-Dale – disant que la réfractivité (à $\lambda_0$) divisée par la masse volumique $\mu$ correspondante est $C(\lambda_0)$ –, donne

$$\tan\psi'_a(\lambda_0)\,\frac{d\psi'_a}{d\lambda_0}(\lambda_0) = \left[\frac{1}{n\left(r_m(\lambda_0),\lambda_0\right)} - \frac{1}{n\left(r_S,\lambda_0\right)}\right]\frac{1}{C(\lambda_0)}\frac{dC}{d\lambda_0}(\lambda_0). \tag{10}$$

Enfin, de l'expression de $\cos\psi'_a(\lambda_0)$, on tire

$$\tan\psi'_a(\lambda_0) = -\sqrt{\frac{1}{\cos^2\psi'_a(\lambda_0)}-1} = -\sqrt{\frac{f^2\left(r_S,\lambda_0\right)-f^2\left(r_m(\lambda_0),\lambda_0\right)}{f^2\left(r_m(\lambda_0),\lambda_0\right)}} \tag{11}$$

et finalement

$$\frac{d\psi'_a}{d\lambda_0}(\lambda_0) = -\frac{r_m(\lambda_0)}{n\left(r_S,\lambda_0\right)}\,\frac{n\left(r_S,\lambda_0\right)-n\left(r_m(\lambda_0),\lambda_0\right)}{\sqrt{r_S^2\,n^2\left(r_S,\lambda_0\right)-r_m^2(\lambda_0)\,n^2\left(r_m(\lambda_0),\lambda_0\right)}}\,\frac{1}{C(\lambda_0)}\frac{dC}{d\lambda_0}(\lambda_0)$$

$$= -\frac{r_m(\lambda_0)}{n\left(r_S,\lambda_0\right)}\,\frac{\dfrac{\partial n}{\partial\lambda_0}\left(r_S,\lambda_0\right)-\dfrac{\partial n}{\partial\lambda_0}\left(r_m(\lambda_0),\lambda_0\right)}{\sqrt{r_S^2\,n^2\left(r_S,\lambda_0\right)-r_m^2(\lambda_0)\,n^2\left(r_m(\lambda_0),\lambda_0\right)}} \tag{12}$$

Or de $\lambda_0 = 656$ nm (raie C) à $\lambda_0 = 486$ nm (raie F), l'intégrale de $\dfrac{1}{C}\dfrac{dC}{d\lambda_0}$ est quasiment le pouvoir dispersif (*i.e.* l'inverse de la constringence), $1/89{,}4$ pour l'air ; on en déduit qu'en radians la valeur absolue de la variation de $\psi'_a$ est généralement petite devant le pouvoir dispersif, mais peut devenir du même ordre que lui si l'observateur se trouve près des frontières du chenal.

**7. Dispersion de la frontière haute d'un chenal à symétrie sphérique**

Elle est beaucoup plus faible, et c'est une des raisons ayant précédemment permis de considérer que $r_m(\lambda_0) \cong r_h$. Mais ici nous voulons préciser ce qui distingue ces deux valeurs, qui ont bien des raisons de présenter théoriquement une petite différence : le maximum de $T(r)$ en $r_h$ ne coïncide pas avec un minimum de $\mu(r)$ ni de $n(r,\lambda_0)$, à cause de la variation de la pression $P(r)$ au voisinage de $r_h$ ; la frontière haute du chenal considéré pour $\lambda_0$ ne correspond pas à un minimum de $n(r,\lambda_0)$ en $r_m(\lambda_0)$, mais de $f(r,\lambda_0)$ – physiquement stationnaire dans l'atmosphère (comme le maximum de la température $T$). Certes, dans la pratique tous ces extrema sont quasiment superposés. Mais pour les distinguer, on part de la propriété rigoureuse de $r_m(\lambda_0)$,

$$\frac{\partial f}{\partial r}\left(r_m(\lambda_0),\lambda_0\right) = 0 = n\left(r_m(\lambda_0),\lambda_0\right)+r_m(\lambda_0)\frac{\partial n}{\partial r}\left(r_m(\lambda_0),\lambda_0\right) \quad \forall\,\lambda_0 \tag{13}$$

dont on déduit la nullité de la dérivée du membre de gauche (fonction de $\lambda_0$ seulement), d'où, après calculs :

$$\frac{dr_m}{d\lambda_0}(\lambda_0) = -\frac{\dfrac{\partial n}{\partial \lambda_0}\big(r_m(\lambda_0), \lambda_0\big) + r_m(\lambda_0)\dfrac{\partial^2 n}{\partial \lambda_0\, \partial r}\big(r_m(\lambda_0), \lambda_0\big)}{r_m(\lambda_0)\dfrac{\partial^2 n}{\partial r^2}\big(r_m(\lambda_0), \lambda_0\big) - 2\dfrac{n\big(r_m(\lambda_0), \lambda_0\big)}{r_m(\lambda_0)}}. \tag{14}$$

De la loi de Gladstone-Dale $n(r, \lambda_0) = 1 + C(\lambda_0)\mu(r)$ qu'on injecte dans (13) et dont on tire les dérivées partielles présentes dans (14), on déduit

$$\frac{dr_m}{d\lambda_0}(\lambda_0) = \frac{1}{C(\lambda_0)\left[r_m^2(\lambda_0)\dfrac{d^2\mu}{dr^2}\big(r_m(\lambda_0)\big) - 2\mu\big(r_m(\lambda_0)\big)\right] - 2}\frac{r_m(\lambda_0)}{C(\lambda_0)}\frac{dC}{d\lambda_0}(\lambda_0). \tag{15}$$

Comme il y a un minimum stationnaire de $\mu(r)$ au voisinage de $r_m(\lambda_0)$, on a $\dfrac{d^2\mu}{dr^2}\big(r_m(\lambda_0)\big) > 0$ qui donne le terme dominant du dénominateur ; ainsi

$$\frac{dr_m}{d\lambda_0}(\lambda_0) \cong \frac{\mu\big(r_m(\lambda_0)\big)}{n\big(r_m(\lambda_0), \lambda_0\big) - 1}\left[r_m(\lambda_0)\frac{d^2\mu}{dr^2}\big(r_m(\lambda_0)\big)\right]^{-1}\frac{1}{C(\lambda_0)}\frac{dC}{d\lambda_0}(\lambda_0). \tag{16}$$

Or $\dfrac{dC}{d\lambda_0}(\lambda_0) < 0$ (dispersion normale pour l'air dans le domaine visible), donc $\dfrac{dr_m}{d\lambda_0}(\lambda_0) < 0$.

*Application numérique*

Dans les cas pratiques, $\mu\big(r_m(\lambda_0)\big)\Big/\dfrac{d^2\mu}{dr^2}\big(r_m(\lambda_0)\big) \sim 10^{-2}\ \text{m}^2$ à $10^1\ \text{m}^2$ ; il n'est guère possible de donner une estimation générale plus précise, tant les situations réelles sont variées, et compliquées par la turbulence. On en déduit seulement que la variation de $r_m$ de la raie F à la raie C est de l'ordre de $-\dfrac{10^{-2}\ \text{m}^2}{(3.10^{-4})\times 6.10^6\ \text{m}}\dfrac{1}{89} \sim -10^{-7}\ \text{m}$ à $-10^{-4}\ \text{m}$. Les effets optiques de cette dispersion sont évidemment indécelables, contrairement à ceux liés à la dispersion beaucoup plus grande de $r_m'$ qui peuvent être spectaculaires, comme nous allons le voir tout de suite.

## 8. Dispersion de la frontière basse d'un chenal à symétrie sphérique

Cette étude s'avère très utile pour interpréter de beaux phénomènes d'optique atmosphérique, où la variation de l'indice de réfraction avec $\lambda_0$ intervient aussi : par exemple, le « rayon vert ». La plus belle de ses formes est celle vue par un observateur situé très près de la frontière basse ($r = r_m'$) du chenal optique (considéré ci-dessus) mais sous elle – celle-ci étant nettement en-dessous de la frontière basse de la couche d'inversion (où $T$ présente un minimum). La dispersion plus importante de $r_m'$ fait que, pour un observateur fixe, la variation de $r_m'(\lambda_0)$ induit un déplacement *perceptible* de cette frontière du chenal (en l'occurrence, on verra que c'est une montée) quand $\lambda_0$ croît – contrairement au déplacement de sa frontière haute. L'aspect de cette forme de « rayon vert » est le plus spectaculaire lorsque l'observateur est juste en-dessous de cette frontière basse pour $\lambda_0$ dans l'indigo (voir la Figure 11), donc encore un peu plus en-dessous dans le vert de façon à donner à la courbe de transfert correspondante un minimum « bien large » mais quand même « net », responsable d'une paire d'images partielles (renversée, et droite accolée juste au-dessus) assez étendue mais nettement détachée de l'image droite inférieure (donc qui est en-dessous) – voir la Figure 11, où les zones assez étendues et quasi triangulaires dans l'ensemble de l'image sont très caractéristiques de cette forme de rayon vert (*i.e.*, vu de sous un chenal).

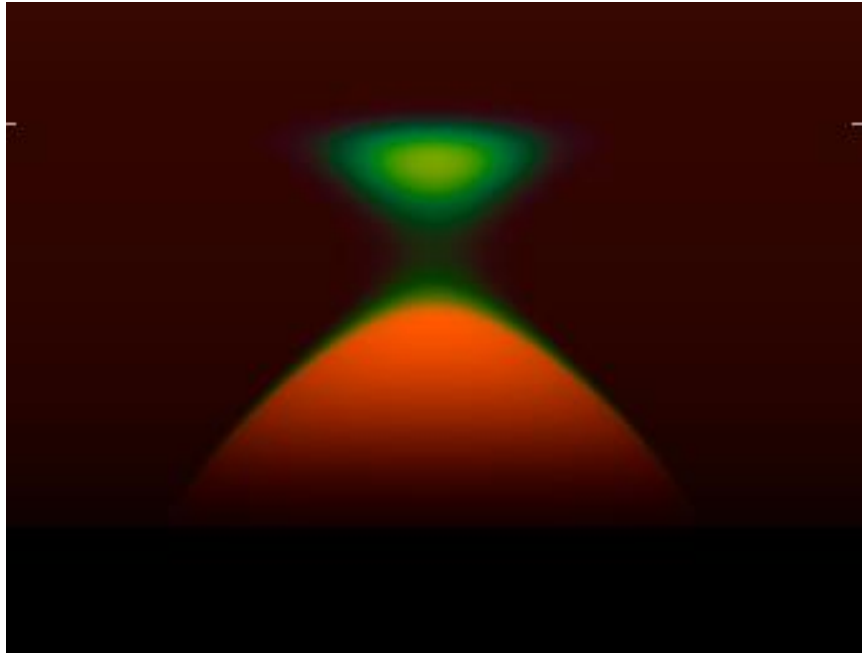

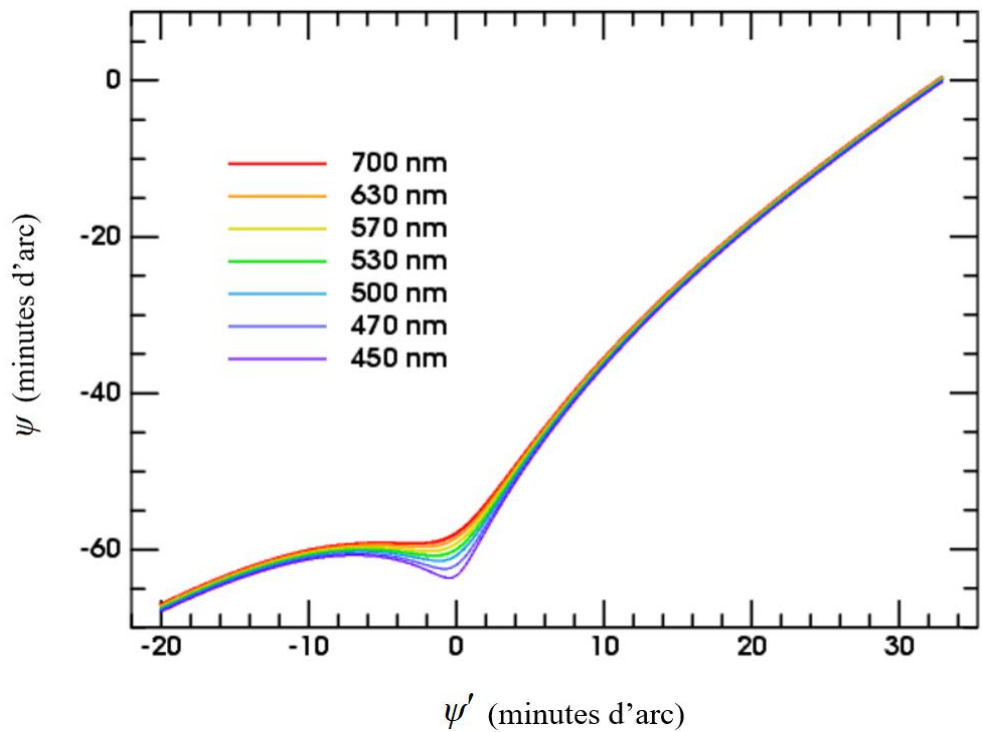

**Figure 11.** Simulation de rayon vert (quand $\psi_u = -60'$) et courbes de transfert correspondantes, pour un observateur situé à 133 m d'altitude, juste sous un chenal pour l'indigo, dû à une très forte couche d'inversion de température s'étendant de 200 à 250 m d'altitude, où la température passe de – 1 °C à 14 °C (une variation totale et monotone de

+ 15 °C est rare mais possible) ; alors le calcul montre que la frontière basse du chenal pour le bleu à $\lambda_0 = 0,48\ \mu m$ se trouve à 134 m d'altitude, et encore un peu moins pour l'indigo à $\lambda_0 = 0,45\ \mu m$ mais quand même supérieure à 133 m, car la courbe de transfert correspondante n'a pas d'asymptote verticale. Pour un objet à distance finie, une courbe de transfert ayant la forme de celle correspondant au rouge (à 700 nm) donnerait une *Fata Morgana*.

© Andrew T. Young, https://aty.sdsu.edu/explain/simulations/sub-duct/SDGF.html

De plus, comme la variation de position des pseudo-images d'un point extra-atmosphérique est un phénomène critique lorsque l'observateur est sur la frontière basse du chenal, les effets chromatiques peuvent être importants et très visibles si la variation d'altitude de cette frontière basse est très grande devant le diamètre de la pupille de l'œil de l'observateur. Dans le but de montrer que c'est bien le cas, il est intéressant d'exprimer $dr'_m/d\lambda_0$. Pour cela, de

$$f\left(r'_m(\lambda_0),\lambda_0\right) = r'_m(\lambda_0)\,n\left(r'_m(\lambda_0),\lambda_0\right) = r_m(\lambda_0)\,n\left(r_m(\lambda_0),\lambda_0\right) = f\left(r_m(\lambda_0),\lambda_0\right) \quad \forall\ \lambda_0 \tag{17}$$

on commence par déduire la relation entre les dérivées logarithmiques de chaque membre (fonction de $\lambda_0$ seulement) :

$$\frac{1}{f\left(r'_m(\lambda_0),\lambda_0\right)}\left[\frac{\partial f}{\partial r}\left(r'_m(\lambda_0),\lambda_0\right)\frac{dr'_m}{d\lambda_0}(\lambda_0) + \frac{\partial f}{\partial \lambda_0}\left(r'_m(\lambda_0),\lambda_0\right)\right] =$$

$$\frac{1}{f\left(r_m(\lambda_0),\lambda_0\right)}\left[\frac{\partial f}{\partial r}\left(r_m(\lambda_0),\lambda_0\right)\frac{dr_m}{d\lambda_0}(\lambda_0) + \frac{\partial f}{\partial \lambda_0}\left(r_m(\lambda_0),\lambda_0\right)\right] = \tag{18}$$

$$\frac{1}{f\left(r_m(\lambda_0),\lambda_0\right)}\frac{\partial f}{\partial \lambda_0}\left(r_m(\lambda_0),\lambda_0\right) = \frac{1}{n\left(r_m(\lambda_0),\lambda_0\right)}\frac{\partial n}{\partial \lambda_0}\left(r_m(\lambda_0),\lambda_0\right)$$

Il est remarquable que $dr'_m/d\lambda_0$ soit découplé de $dr_m/d\lambda_0$ ; cela est dû à la stationnarité de $f$ en $r_m$. De la loi de Gladstone-Dale on tire, par dérivation logarithmique,

$$\frac{1}{n\left(r'_m(\lambda_0),\lambda_0\right)-1}\frac{\partial n}{\partial \lambda_0}\left(r'_m(\lambda_0),\lambda_0\right) = \frac{1}{C(\lambda_0)}\frac{dC}{d\lambda_0}(\lambda_0) = \frac{1}{n\left(r_m(\lambda_0),\lambda_0\right)-1}\frac{\partial n}{\partial \lambda_0}\left(r_m(\lambda_0),\lambda_0\right) \tag{19}$$

donc, en combinant les deux relations précédentes,

$$\frac{1}{f\left(r'_m(\lambda_0),\lambda_0\right)}\frac{\partial f}{\partial r}\left(r'_m(\lambda_0),\lambda_0\right)\frac{dr'_m}{d\lambda_0}(\lambda_0) = \frac{1}{n\left(r_m(\lambda_0),\lambda_0\right)}\frac{\partial n}{\partial \lambda_0}\left(r_m(\lambda_0),\lambda_0\right)$$

$$-\frac{1}{n\left(r'_m(\lambda_0),\lambda_0\right)}\frac{\partial n}{\partial \lambda_0}\left(r'_m(\lambda_0),\lambda_0\right) = \left[\frac{n\left(r_m(\lambda_0),\lambda_0\right)-1}{n\left(r_m(\lambda_0),\lambda_0\right)} - \frac{n\left(r'_m(\lambda_0),\lambda_0\right)-1}{n\left(r'_m(\lambda_0),\lambda_0\right)}\right]\frac{1}{C(\lambda_0)}\frac{dC}{d\lambda_0}(\lambda_0) \tag{20}$$

Enfin, de la relation (43) de l'article [5], on déduit que

$$\frac{1}{f\left(r'_m(\lambda_0),\lambda_0\right)}\frac{\partial f}{\partial r}\left(r'_m(\lambda_0),\lambda_0\right) = \frac{1-\kappa\left(r'_m(\lambda_0),\lambda_0\right)}{r'_m(\lambda_0)} \tag{21}$$

où $\kappa\left(r'_m(\lambda_0),\lambda_0\right)$ désigne le coefficient de réfraction de l'atmosphère en $r'_m(\lambda_0)$ pour $\lambda_0$ (défini au sous-paragraphe 7.1.1 de [5]). Ainsi,

$$\frac{1-\kappa\left(r'_m(\lambda_0),\lambda_0\right)}{r'_m(\lambda_0)}\frac{dr'_m}{d\lambda_0}(\lambda_0) = \frac{1}{n\left(r'_m(\lambda_0),\lambda_0\right)}\left[1-\frac{r_m(\lambda_0)}{r'_m(\lambda_0)}\right]\frac{1}{C(\lambda_0)}\frac{dC}{d\lambda_0}(\lambda_0) \tag{22}$$

grâce à la relation (17), et finalement

$$\frac{dr'_m}{d\lambda_0}(\lambda_0) = \frac{-1}{1-\kappa\left(r'_m(\lambda_0),\lambda_0\right)}\frac{r_m(\lambda_0)-r'_m(\lambda_0)}{n\left(r'_m(\lambda_0),\lambda_0\right)}\frac{1}{C(\lambda_0)}\frac{dC}{d\lambda_0}(\lambda_0). \tag{23}$$

Comme $r_m(\lambda_0) - r'_m(\lambda_0) > 0$ dans le cas d'un chenal dû à une couche d'inversion raide, que $1 - \kappa\left(r'_m(\lambda_0), \lambda_0\right) > 0$ si on voit le Soleil sur l'horizon astronomique, et que $\dfrac{dC}{d\lambda_0}(\lambda_0) < 0$ en dispersion normale, on a $\dfrac{dr'_m}{d\lambda_0}(\lambda_0) > 0$, contrairement à $\dfrac{dr_m}{d\lambda_0}(\lambda_0)$. Les déplacements chromatiques des frontières haute et basse du chenal sont en sens inverse ; la largeur de celui-ci est moins grande dans le rouge que dans le bleu.

*Application numérique*
Avec la couche d'inversion très forte et raide considérée pour la Figure 11, qui commence à 200 m d'altitude (correspondant à $r_b$) et finit à 250 m (correspondant à $r_h \cong r_m(\lambda_0)$), où $T$ augmente de 15 °C de bas en haut (valeur typique des conditions de Santa Ana sur la côte californienne), on a (pour $\lambda_0$ dans le bleu) un chenal optique commençant à 134 m d'altitude (qui correspond à $r'_m$) ; ainsi $r_m(\lambda_0) - r'_m(\lambda_0) \cong 116$ m. En prenant la valeur standard $\kappa \cong 1/5,99$ à 134 m d'altitude, on en déduit que la variation de $r'_m$ de la raie F à la raie C est voisine de $\dfrac{1}{1-(1/5,99)}\,\dfrac{250\text{ m} - 134\text{ m}}{1,0003}\,\dfrac{1}{89,4} \cong 1,56$ m. Cette valeur est assez grande pour donner des effets chromatiques bien visibles, par exemple sur une photo prise avec un téléobjectif – son diamètre étant très inférieur à 1,56 m.

## 9. Conclusion

Cette étude donne un aperçu de la puissance de la discussion par quadrature pour l'étude analytique de la réfraction astronomique à symétrie sphérique ; même s'il s'agit d'un cas idéalisé, c'est celui qui est utilisé pour construire les tables de réfraction [15]. Pour une utilisation informatique, une quadrature présente une difficulté (la question du signe de la dérivée première), alors on préfère étudier numériquement l'équation différentielle équivalente, du second ordre, pour laquelle il n'y a plus de signe à discuter [16] ; elle se déduit de la quadrature par une procédure très classique, décrite dans notre article [5] dévolu à la réfraction astronomique à symétrie sphérique dont nous présentons encore bien d'autres propriétés.

**Références**

[1] V. Arnold, *équations différentielles ordinaires*, 4e éd., Mir, Moscou, 1988, voir p. 132-145.
[2] A. T. Young, G. W. Kattawar, « Sunset science. II. A useful diagram », *Appl. Opt.* **37** (1998), p. 3785-3792.
[3] L. Dettwiller, « L'invariant de Bouguer et ses conséquences : commentaire historique », *C. R. Phys.* **23** (2022), p. 415-452.
[4] L. Dettwiller, « La discussion par Kummer d'une quadrature sur la réfraction astronomique : commentaire historique », *C. R. Phys.* **23** (2022), p. 503-525.
[5] L. Dettwiller, « Propriétés remarquables de la réfraction astronomique dans une atmosphère à symétrie sphérique », *C. R. Phys.* **23** (2022), p. 63-102.
[6] J. D. Everett, « On the optics of mirage », Phil. Mag. **45** (1873), p. 161-172, voir p. 169.

[7] J.-P. Pérez, *Optique géométrique et ondulatoire*, 4ᵉ éd., Masson, Paris, 1994, voir p. 169.

[8] M. Born, E. Wolf, *Principles of Optics – Electromagnetic Theory of Propagation, Interference and Diffraction of Light*, 6th ed., Pergamon Press, Oxford, 1980, voir p. 124.

[9] E. E. Kummer, « Über atmosphärische Strahlenbrechung », *Monatsber. Kgl. Preuss. Akad. Wiss. Berlin* **5** (1860), p. 405-420 – traduit par Verdet dans *Ann. Chim. Phys.*, série 3, **61** (1861), p. 496–507.

[10] J. L. Dauvergne. (Page consultée le 9 janvier 2022), *Coucher de Soleil en temps réel au Pic du Midi* [En ligne]. Adresse URL : https://www.youtube.com/watch?v=GwIphktocoQ

[11] J. Cassini, *Traité de la grandeur et de la figure de la Terre*, Pierre de Coup, Amsterdam, 1723, p. 141.

[12] A. T. Young, G. W. Kattawar, P. Parviainen, « Sunset science. I. The mock mirage », *Appl. Opt.* **36** (1997), p. 2689-2700.

[13] L. Dettwiller, « Phénomènes de réfraction atmosphérique terrestre », *C. R. Phys.* **23** (2022), p. 103-132.

[14] A. Wegener, « Elementare Theorie der atmosphärischen Spiegelungen », *Annalen der Physik*, series 4, **57** (1918)**,** p. 203-230.

[15] F. Mignard, « Les tables de réfraction astronomique », *C. R. Phys.* **23** (2022), p. 133-178.

[16] F. Mignard, V. Brumberg, « Corrections pour la réduction des observations optiques », in *Introduction aux éphémérides et phénomènes astronomiques – Supplément explicatif à la Connaissance des Temps* (J. Berthier, P. Descamps, F. Mignard, éds.), IMCCE / EDP Sciences, Paris, 2021, p. 597-643.